\documentclass[conference]{IEEEtran}
\IEEEoverridecommandlockouts
\usepackage{cite}
\usepackage{overpic}
\usepackage{amsmath,amssymb,amsfonts}
\usepackage{graphicx}
\usepackage{textcomp}
\usepackage{xcolor}
\usepackage{float}
\usepackage{amsthm}
\usepackage{graphicx}
\usepackage{epstopdf}
\usepackage{subfigure}
\usepackage{adjustbox}
\usepackage{stfloats}
\usepackage{amsmath,bm}
\usepackage{amsfonts}
\usepackage{amssymb}
\usepackage{color}
\usepackage{multirow}
\usepackage{multicol}
\usepackage{soul,xcolor}
\usepackage{algorithm}
\usepackage{algpseudocode}%
\usepackage{enumitem}
\usepackage{booktabs} 

\usepackage{url}

\theoremstyle{plain}

\newcommand{\vect}[1]{\mathbf{#1}}

\def\Ttran{\mbox{\tiny $\mathrm{T}$}}

\begin{document}

\title{EARL: Energy-Aware Adaptive Antenna Control with Reinforcement Learning in O-RAN Cell-Free Massive MIMO Networks
\thanks{This work has been part of 6G-SUSTAIN: Sensing Integrated Elastic 6G Networks for Sustainability project funded by Vinnova in Sweden, and partially funded by the project ``Celtic-Next project RAI-6Green: Robust and AI Native 6G for Green Networks'' with project-id: C2023/1-9 also by Vinnova in Sweden. This work was supported by the Swedish Innovation Agency  (VINNOVA) through the SweWIN center (2023-00572).}
}

\author{\IEEEauthorblockN{Zilin Ge\IEEEauthorrefmark{1}, Ozan Alp Topal\IEEEauthorrefmark{1}, Irshad Ahmad Meer\IEEEauthorrefmark{1}, Pei Xiao\IEEEauthorrefmark{2}, and Cicek Cavdar\IEEEauthorrefmark{1}}
\IEEEauthorblockA{ \IEEEauthorrefmark{1}{Department of Communication Systems, KTH Royal Institute of Technology, Stockholm, Sweden
		}   \\
        \IEEEauthorrefmark{2}{ Institute for Communication Systems (ICS), University of Surrey, Guildford, United Kingdom
		} \\
		\IEEEauthorblockA{E-mail: 
        \{zilin, oatopal, iameer, cavdar\}@kth.se,  p.xiao@surrey.ac.uk}
}

}

\maketitle

\begin{abstract}
Cell-free massive multi-input multi-output (MIMO) promises uniform high performance across the network, but also brings a high energy cost due to joint transmission from distributed radio units (RUs) and centralized processing in the cloud. Leveraging resource-sharing capabilities of Open Radio Access Network (O-RAN), we propose EARL, an energy-aware adaptive antenna control framework using reinforcement learning. EARL dynamically configures antenna elements in RUs to minimize radio, optical fronthaul, and cloud processing power consumption while meeting user spectral efficiency demands. Numerical results show power savings of up to $81\%$ and $50\%$ over full-on and heuristic baselines, respectively. The RL-based approach operates within $220$ ms, satisfying O-RAN’s near-real-time limit, and a greedy refinement further halves power consumption at a $2$ s runtime.
\end{abstract}


\section{Introduction}

Cell-free massive multi-input-multi-output (MIMO) offers a uniform user experience through coherent joint transmission and reception by densely deployed radio units (RUs) connected to a central cloud. 
The Open Radio Access Network (O-RAN) provides the key architectural enablers for a scalable cell-free massive MIMO deployment, including baseband virtualization and dynamic resource sharing across the cloud and fronthaul domains \cite{ozlem_jsac}. 
Despite these advantages, dense O-RU and fronthaul deployment increases both deployment and operational costs. O-RAN’s resource-sharing capability allows dynamic deactivation of idle RUs, processing units, and fronthaul links, enabling up to $35\%$ end-to-end power savings through joint RU activation and power allocation \cite{ozlem_jsac}. 
Further granularity at the antenna-chain level activation provides up to $40\%$ higher energy savings \cite{ozan_asilomar}. However, existing methods face two key challenges.

First, they rely on optimization-based approaches, which are computationally intensive and impractical for real-time deployment. On the other hand, machine learning, particularly reinforcement learning (RL), offers faster decision-making for resource allocation. Deep RL (DRL) has been applied to downlink power control and RU shutdown under spectral efficiency (SE) constraints \cite{luo_power_2022, wu_energy-efficient_2024}. Multi-agent DRL schemes enable distributed RU activation, where each RU acts independently \cite{topal_drl-based_2023}. However, these methods neglect antenna-level control and ignore cloud processing and fronthaul power consumption, limiting achievable energy savings. Second, the enlisted works assume distributed precoding, where each RU performs interference cancellation locally, leading to suboptimal SE and increased RU activation. 
In contrast, O-RAN split options \text{8} and \text{7.1}, as defined by 3GPP \cite{3gpp_tr_38_816}, enable centralized precoding and channel estimation in the O-Cloud, providing significant SE gains, especially for cell-edge users, as shown in Figs. 6.3 and 6.5 in \cite{cell-free-book}. However, an energy-efficient antenna or RU configuration has not been solved considering centralized precoding.

To address these limitations, we propose EARL, an Energy-Aware Reinforcement Learning framework, for adaptive antenna and O-RU control under centralized precoding. EARL functions as an energy-aware xApp in the O-Cloud, dynamically allocating antennas, RUs, and fronthaul resources based on long-term channel characteristics to minimize total power consumption while satisfying UE SE requirements. The main contributions of this work are as follows:

\begin{itemize}
\item We formulate an end-to-end power consumption model for cell-free massive MIMO under O-RAN split options 8 and 7.1, predominantly scaling with active O-RUs and antenna ports.
\item We cast the SE-constrained power minimization problem into an RL framework with penalized constraint violations and develop a scalable heuristic for comparison. 
\item EARL achieves up to $81\%$ and $50\%$ energy savings over full-on and heuristic schemes, respectively. 
The RL-only version operates within $220$ ms, well below the O-RAN near-real-time limit, while a greedy refinement further halves the power consumption with a $2$ s runtime.
\end{itemize}

\section{System Model}
We consider the downlink of a cell-free massive MIMO system built on an O-RAN architecture operating in time-division duplex (TDD) mode and orthogonal frequency-division multiplexing (OFDM). The system consists of $L$ O-RUs connected to a central cloud unit with virtualization and processing resource sharing capabilities, via lossless fiber fronthaul connections \cite{ref10}. 
Each O-RU is equipped with $N$ antennas, and the O-RUs serve $K$ single-antenna UEs over a mid-band frequency (sub-6 GHz) channel. 

\subsection{Channel Model}
 We assume correlated Rayleigh fading channels, i.e., the channel from UE $k$ to O-RU $l$ is $\vect{h}_{kl} \sim \mathcal{N}_{\mathbb{C}}(\vect{0}, \vect{R}_{kl})$. $\vect{R}_{kl} \in \mathbb{C}^{N \times N}$ is the correlation matrix, determined by the spatial correlation of the channel $\vect{h}_{kl}$ between the antennas of O-RU $l$ and the
corresponding average channel gain, which is denoted by $\beta_{kl} = \operatorname{tr}(\vect{R}_{kl})/N$. 

We consider that each O-RU can activate or deactivate its antenna ports based on the decision given by EARL. Therefore, we denote the activated antennas at O-RU $l$  by $n_l \in \{0,\ldots,N\}$, where $n_l=0$ means that RU is deactivated. 

We let $\tau_c$ denote the number of symbols in a TDD frame, where it consists of uplink training with $\tau_p$ symbols, and downlink data transmission with $\tau_d=\tau_c-\tau_p$ symbols. We assume that during the channel estimation phase, all antennas in all  O-RUs are active \cite{ericsson_2022_energy_5g}. Applying minimum mean-squared error (MMSE) channel estimation, the estimated channel vector between O-RU $l$ and UE $k$ is denoted by $\hat{\vect{h}}_{kl} \sim \mathcal{N}_{\mathbb{C}}\left(\mathbf{0}, \vect{B}_{kl} \right)$ and the channel estimation error $\tilde{\mathbf{h}}_{k l} \sim \mathcal{N}_{\mathbb{C}}\left(\mathbf{0}, \mathbf{C}_{k l}\right)$, where $\vect{B}_{kl}$ and $\mathbf{C}_{k l}$ denote the estimate and estimation error correlation matrix, respectively \cite[Chapter~4]{cell-free-book}.

\subsection{Downlink Data Transmission}
\label{sec:centralized_downlink}
Different from \cite{ozlem_jsac, ozan_asilomar}, we consider centralized downlink operation in this work. From the network architecture perspective, precoding operation must be carried out in the O-Cloud, allowing for functional split options 8 and 7.1.  The received downlink signal at UE $k$ is
\begin{equation}
    y_k = 
    \mathbf{h}_k^{H} \mathbf{D}_k \mathbf{w}_k \varsigma_k
    + \sum_{i \neq k} \mathbf{h}_k^{H} \mathbf{D}_i \mathbf{w}_i \varsigma_i
    + \varrho_k ,
    \label{eq:dl_received}
\end{equation}
where $\mathbf{h}_k \in \mathbb{C}^{LN}$ denotes the collective channel from all $L$ O-RUs to UE $k$, as $\mathbf{h}_k = [\mathbf{h}^{\Ttran}_{k1}, \ldots, \mathbf{h}^{\Ttran}_{kL}]^{\Ttran}$. $\mathbf{w}_i = \left[\mathbf{w}_{i 1}^{\Ttran} \ldots \mathbf{w}_{i L}^{\Ttran}\right]^{\Ttran} \in \mathbb{C}^{LN}$ is the precoding vector for UE $i$. Note that, if an O-RU does not serve a UE, we assume $\mathbf{w}_{i l} = \vect{0}$. $\varsigma_i$ is the unit-power data symbol, and $\varrho_k \sim \mathcal{N}_{\mathbb{C}}(0, \sigma^2)$ is additive white Gaussian noise. The matrix $\mathbf{D}_i \in \{0,1\}^{LN\times LN}$ is a block-diagonal matrix, which jointly determines the activation of the antennas as an O-RU and UE association, where $\mathbf{D}_i = \operatorname{diag}(\vect{D}_{i1}, \ldots, \vect{D}_{iL})$, and the number of active antennas at O-RU $l$ can be given by
\begin{equation}
    n_l = \sum_{l=1}^L \operatorname{Tr}\left( \hat{\vect{D}}_l \right),
\end{equation}
where $\hat{\vect{D}}_l$ is composed of the element wise maximum of association matrices as $\hat{\vect{D}}_l[q,w] = \max_{i=1,\ldots, K}(\vect{D}_{il}[q,w]), \, \forall q,w \in \{1, \ldots, N\}$.
For tractability, we assume that all active O-RUs are simultaneously serving all UEs. 
As given in Theorem 6.1 of \cite{cell-free-book}, an achievable downlink spectral efficiency (SE) for UE $k$ is
\begin{equation}
    \mathrm{SE}_k
    = \frac{(\tau_c-\tau_p)}{\tau_c}
      \log_2 \!\left(1 + \mathrm{SINR}_k \right)
      \quad \text{[bit/s/Hz]},
    \label{eq:SE_dl}
\end{equation}
and the effective downlink SINR is
\begin{equation}
    \mathrm{SINR}_k
    =
    \frac{\left|\mathbb{E}\{\mathbf{h}_k^{H} \mathbf{D}_k \mathbf{w}_k \right\}|^2}
    { \sum_{i=1}^{K}
        \mathbb{E}\{ \left| \mathbf{h}_k^{H} \mathbf{D}_i \mathbf{w}_i \right|^2  \}
        - \left |\mathbb{E}\{\mathbf{h}_k^{H} \mathbf{D}_k \mathbf{w}_k \} \right|^2
        + \sigma^2 } .
    \label{eq:SINR_dl}
\end{equation}
By fully benefiting from the centralized processing capability, we consider minimum mean-square error (MMSE) precoding, that is obtained from
\begin{equation}
 \bar{\mathbf{w}}_k
    = p_k
    \left(
        \sum_{i =1}^K
        p_i
        \mathbf{D}_k (\hat{\mathbf{h}}_i \hat{\mathbf{h}}_i^{H} + \vect{C}_i) \mathbf{D}_k
        + \sigma^2 \mathbf{I}_{LN}
    \right)^{-1}
    \mathbf{D}_k \hat{\mathbf{h}}_k ,
    \label{eq:pmmse_precoder}
\end{equation}
where $ \mathbf{w}_k = \frac{ \sqrt{\rho_k}\bar{\mathbf{w}}_k}{\sqrt{\mathbb{E}(\|\bar{\mathbf{w}}_k\|^2)}}$ is the precoding vector. $p_k$ and $\rho_k$ are the uplink and downlink power assigned for UE $k$, respectively. Note that the proposed methodology can work with other linear precoding schemes as well, by adapting $\mathbf{w}_k$ accordingly. To lower the dimension of the action space, we adopt an efficient power allocation scheme in Eq. (7.3) of~\cite{cell-free-book}, originally formulated as
\begin{equation}
    \rho_k =
    \rho_{\max}
    \frac{
        \Bigl(\sum_{l = 1}^L \beta_{kl}\Bigr)^{\upsilon}
        \omega_k^{-\kappa}
    }{
        \displaystyle
        \max_{l = 1, \ldots, L}
        \sum_{i = 1}^K
        \Bigl(\sum_{l = 1}^L \beta_{il}\Bigr)^{\upsilon}
        \omega_i^{1-\kappa}
    },
    \label{eq:rho_fractional}
\end{equation}
where $\omega_k= \max_l\{\mathbb{E}(\|\bar{\vect{w}}_{kl}\|)\}$.  $\rho_k$ denotes the total downlink transmit power allocated to UE~$k$, $\rho_{\max}$ is the maximum allowed transmit power per O-RU. $\upsilon \in [-1,1]$ and $\kappa \in [0,1]$ are tuning coefficients controlling the trade-off between fairness and sum throughput.

\section{End-to-End Power Minimization}
In this section, we first explain the end-to-end power consumption model considered, and then we formulate the power minimization problem. 

\subsection{Power Consumption Model}
The total power consumption of the considered cell-free massive MIMO system with optical fronthaul includes the contributions from all O-RUs, the optical fronthaul, and the centralized cloud unit \cite{ozlem_jsac}. Accordingly, the total power can be given by
\begin{equation}
    P_{\text{tot}} =
    \sum_{l=1}^{L} \left( P_{\text{RU}, l} + P_{\text{fh}, l} \right)
    + P_{\text{cloud}} + P_{\text{fh,cloud}},
    \label{eq:ptot}
\end{equation}
where $P_{\text{RU},l}$  denotes the sum  radio and processing power at O-RU $l$, $P_{\text{fh},l}$ denotes the fronthaul power at O-RU $l$.   $P_{\text{cloud}}$, $P_{\text{fh,cloud}}$ represent the cloud-site baseband processing and fronthaul power, respectively.

In this work, we consider Options 8 and 7.1 with centralized cell-free operation. Based on the chosen functional split, the processing-related power consumption will shift between O-RU and O-Cloud.

\subsubsection{O-RU-Site Power Consumption}
Power consumption at O-RU site is caused by two main factors: $(1)$ transmit and hardware power consumption; $(2)$ power consumption for processing performed at the RU-site (depends on the chosen functional split). The power consumption of O-RU $l$  becomes 
\begin{equation}
    P_{\text{RU},l} =
    n_l P_{\text{st}}
    + \Delta^{\rm tr} \sum_{k=1}^K \rho_k + P^{\mathrm{proc}}_{\mathrm{RU},l},
\end{equation}
where $n_l$ is the number of active antennas at O-RU~$l$, $P_{\text{st}}$ is the static power consumption per active RF chain,  $\Delta^{\rm tr}\geq 1$ is the slope of the load-dependent transmit power consumption. $P^{\mathrm{proc}}_{\mathrm{RU},l}$ is the power consumption by the processing performed at O-RU $l$, and it depends on the chosen functional split, and is calculated by  
\begin{equation}
P^{\mathrm{proc}}_{\mathrm{RU},l} = \frac{\mathcal{X}}{{\sigma_{\text{cool}}}} \left( P^{\mathrm{proc}}_{\mathrm{RU},0} +   \frac{ C_{\mathrm{RU},l} }{ C_{\mathrm{RU},\mathrm{max}}}\right),
\end{equation}
where $P^{\mathrm{proc}}_{\mathrm{RU},0}$ is the idle processing power, and $C_{\mathrm{RU},l}$ is the giga operations per second (GOPS) at the O-RU site. The value of $C_{\mathrm{RU},l}$ can be calculated by summing the processes given in Table \ref{tab:GOPS_table} marked by O-RU for the chosen functional split. $\mathcal{X}$ is a binary variable that is equal to one for Option 7.1, and zero for Option 8. $ C_{\mathrm{RU},\mathrm{max}}$ is the processor efficiency at RU in terms of GOPS/W, which can vary based on the chosen hardware technology. $0 < \sigma^{\mathrm{RU}}_{\mathrm{c}} \leq 1$ is the cooling efficiency at any RU.

\subsubsection{O-Cloud-Site Power Consumption}
O-Cloud performs the high-physical-layer baseband operations, including precoding, modulation, and coding. Its power consumption is modeled as
\begin{equation}
    P_{\text{cloud}} =
    P_{\text{fixed}}
    + \frac{1}{\sigma_{\text{cool}}}
      \left(
      P_{\text{idle,cloud}}
      + \frac{ C_{\mathrm{Cloud}} }{ C_{\mathrm{Cloud},\mathrm{max}} }
      \right),
\end{equation}
where $P_{\text{fixed}}$ is the load-independent baseline power, 
$P_{\text{idle,cloud}}$ is the idle processing power, 
$C_{\text{cloud}}$ represents the total computational load (in GOPS) of the centralized PHY functions. $C_{\mathrm{Cloud},\mathrm{max}}$ is the processor efficiency at vDU/vCU-site in terms of GOPS/W, where based on the chosen hardware technology. $\sigma_{\text{cool}}$ models data-center cooling efficiency.

\subsubsection{Optical Fronthaul Power Consumption}
The optical fronthaul includes both O-RU and O-Cloud-side components. 
At each O-RU, an optical transceiver consumes a constant power whenever the O-RU is active:
\begin{equation}
    P_{\text{fh},l} =
    P_{\text{opt}}\, \mathbb{I}_{\{n_l>0\}},
\end{equation}
where $P_{\text{opt}}$ is the per-O-RU optical transceiver power and $\mathbb{I}_{\{n_l>0\}}$ equals~1 if O-RU~$l$ is active. 
At the cloud, the fronthaul power grows with the aggregate traffic load:
\begin{equation}
    P_{\text{fh,cloud}} =   {\sigma^{-1}_{\text{cool}}}\,\varsigma_{\text{fh}}\, C_{\text{fh}},
\end{equation}
where $\varsigma_{\text{fh}}$ converts traffic load $C_{\text{fh}}$ to power consumption \cite{ozlem_jsac}.

\subsection{GOPS Analysis}
 In this work, we consider Option 7.1 and Option 8 as possible functional splits. In Option 7.1, the radio frequency (RF) layer operations and low-PHY layer processing are carried out at the O-RU, while higher PHY processes, such as modulation and coding, are carried out at the O-Cloud. In Option 8, all processing is carried out at the O-Cloud. The GOPS for the operations considered in this work are given in Table \ref{tab:GOPS_table} taken from \cite{ozlem_jsac}. The factors demonstrate that most of the processing scales with the active number of antenna ports.  $\mathrm{W}_r$ and $\mathrm{SE}_r$ denote the ratio of the bandwidth and the ratio of the SE of a UE for this work to the reference setup \cite{ozlem_jsac}. In the reference setup, $20$\,MHz bandwidth is chosen, and the SE is equal to $6$\,bit/s/Hz. The binary variable $r_{il}$ takes the value of $1$ if O-RU $l$ serves UE $i$ and zero otherwise. $T_s$ is the OFDM symbol duration, $N_{\rm DFT}$ is the DFT size, $N_{\rm used}$  is the number of used subcarriers, and  $f_s$ is the sampling rate.

\begin{table}[tb]
    \centering
    \caption{GOPS formulations for various operations.}
    \begin{tabular}{l|l|l|l|l}
    Function & GOPS per unit* & Factor & 8 & 7.1 \\ \hline
    $C_{\mathrm{filter}, l}$ & ${40 f_s}/{10^9}$ & $n_l$ & \multirow{7}{*}{O-Cl} &  \\ 
    $C_{\mathrm{DFT}, l}$ & $\frac{8 N_{\mathrm{DFT}} \log_2(N_{\mathrm{DFT}})}{T_s10^9}$ & $n_l$ &  & O-RU \\ 
    $C_{\mathrm{map},l}$ & $1.3 \mathrm{W}_r \mathrm{SE}^{1.5}_r$ & $\sum_{i=1}^K r_{il}$ &  &  \\
    \cline{5-5}
    $C_{\mathrm{prec}, l}$ & $\left(\frac{8 \tau_d N_{\mathrm{used}}}{T_s 10^9 \tau_c} \right)$ & $n_l \sum_{i=1}^K r_{il}$ &  &  \\    
    $C_{\mathrm{mod},l}$ & $1.3 \mathrm{W}_r$ & $n_l$ &  & O-Cl \\
    $C_{\mathrm{cod},l}$ & $5.2 \mathrm{W}_r \mathrm{SE}_r$ & $\sum_{i=1}^K r_{il}$ &  &  \\
    $C_{\mathrm{netw},l}$ & $8 \mathrm{W}_r \mathrm{SE}_r$ & $1$  &  &  \\ 
    \hline
    \end{tabular}
    
    \footnotesize{  \vspace{2mm} *Total GOPS calculated by multiplying GOPS per unit and unit factor.}
    
    \label{tab:GOPS_table}
\end{table}

\subsection{Problem Statement}
The total power consumption $P_{\text{tot}}$ in
\eqref{eq:ptot}, explicitly depends on $\mathbf{n} = [n_1, \ldots, n_L]$ through both the hardware and fronthaul components. As shown in \cite{ozan_asilomar}, the main factor affecting the total power consumption is the active number of antenna elements, and then the active number of O-RUs. Therefore, we removed the terms related to other control variables from $P_{\text{tot}}$, and formulated the problem as 
\begin{subequations} \label{eq:power_min_opt0:problem}
\begin{align}
 & \underset{\{\vect{n}\}}{\text{minimize}} \quad  c_0 \|\vect{n}\|_0 + c_1 \|\vect{n}\|_1 \label{eq:power_min_opt0:objective} \\
     & \textrm{subject to} \nonumber \\ &\operatorname{SINR}_k\geq \upsilon_k, \quad \forall k \label{eq:power_min_opt0:SINR_constraint} \\
& n_l \in \{0,\ldots, N\}, \quad \forall l \label{eq:power_min_opt0:integer} .
\end{align}
\end{subequations}
The objective function, \eqref{eq:power_min_opt0:objective}, aims to minimize the end-to-end total power consumption by minimizing the number of active antenna elements and the number of active O-RUs, where 
\begin{align}
    c_0 &= \frac{C_{\mathrm{netw},l} + K C_{\mathrm{cod},l} }{C_{\mathrm{cloud, max}}\sigma_{\mathrm{cool}}}+ P^{\mathrm{proc}}_{0} + P_{\text{opt}}, \\
    c_1 &= P_{\mathrm{st}} +  \frac{ C_{\mathrm{mod},l} + K C_{\mathrm{prec},l} + (1-\mathcal{X})(C_{\mathrm{filter},l} + C_{\mathrm{DFT},l})  }{C_{\mathrm{GPP}}^{\max }\sigma_{\mathrm{cool }}} \nonumber \\ &+ \frac{1 }{C_{\mathrm{RU, max}}} \left[ \mathcal{X}(C_{\mathrm{filter},l} + C_{\mathrm{DFT},l}) \right].
\end{align}
\eqref{eq:power_min_opt0:SINR_constraint} ensures that effective SINR at UE $k$ is higher than or equal to the threshold value $\upsilon_k$. \eqref{eq:power_min_opt0:integer} ensures that the number of active antennas at O-RU $l$, $n_l$, is an integer variable smaller than or equal to the deployed number of antennas at O-RU $l$. This problem is non-convex and of a combinatorial nature due to the zero-norm term in the objective function, the non-convex expressions in \eqref{eq:power_min_opt0:SINR_constraint}, and integer variables defined in \eqref{eq:power_min_opt0:integer}. 
The main challenge in solving this problem with optimization tools is that the effective SINR for any user cannot be expressed as a function of $\vect{n}$, the active number of antennas, when centralized precoding is considered. Therefore, in the following, we will reformulate this problem under the RL framework.

\section{Reinforcement Learning based Antenna Activation Framework}
\label{sec:rl_gr}
In this section, we present an RL-based algorithm to solve \eqref{eq:power_min_opt0:problem}. The trained RL agent can be implemented as an energy-saving xAPP, located in the near-RT RIC. Since we consider O-DU and O-CU are bundled (named as O-Cloud), near-RT RIC, and consequently, the proposed RL agent lies in the O-Cloud. The agent requires only the path loss information of the UEs in the network, and receives it from the E2 interface \cite{oran_alliance}. Later, the agent should decide on the activation of RF chains in O-RU, and should propagate its decision again with the E2 interface to the network within 1 second. We assume that through other network controllers, the fronthaul load, the radio processing will also be scaled intelligently, lowering the total energy consumption \cite{ozlem_jsac}. Below, we provide the action, the state spaces, and the reward function. 

\textbf{Action space:} 
At each decision epoch $t$, the near-RT controller produces an incremental action vector 
$\boldsymbol{a}_t \!\in\! \{-1,0,+1\}^L$, where $a_{l,t}$ denotes the configuration in the number of 
active antenna elements at O-RU~$l$ equipped with $K$ antennas in total. 
The configuration evolves as $n_{l,t+1}=\mathrm{clip}(n_{l,t}+a_{l,t},\,0,\,N)$, ensuring \eqref{eq:power_min_opt0:integer}.

\textbf{State space:}
We let $\boldsymbol{\Phi}\!\in\!\mathbb{R}^{L\times K}$ denote the channel gain matrix, where $\boldsymbol{\Phi}[l, k] = \beta_{lk}$. 
At decision epoch~$t$, the system state is defined as
\[
s_t = \big[\, \mathrm{vec}(\boldsymbol{\Phi}),\, \mathbf{n}_t,\, \frac{P_{\mathrm{tot}}}{P^{\mathrm{max}}_{\mathrm{tot}}},\, R_{\mathrm{vio}}(t) \,\big],
\]
where $\mathbf{n}_t = [n_{1t}, \ldots, n_{Lt}]$ captures the current antenna element activation status of all O-RUs, 
while $P_{\mathrm{tot}}$ is normalized by the maximum power consumption $P^{\mathrm{max}}_{\mathrm{tot}}$.  $R_{\mathrm{vio}}(t)=\frac{1}{K}\sum_{k=1}^{K}\mathbb{I}\{\mathrm{SE}_k<\Gamma^{\min}\}$ 
quantifies the fraction of users violating the SE constraint, where $\Gamma^{\min}$ is the minimum rate requirement.
This ensures that the RL agent observes both the network conditions and 
its present configuration before determining the next incremental action $\boldsymbol{a}_t$.

\textbf{Reward function:}
The reward encourages energy efficiency while ensuring users' QoS requirements. 
Each UE must achieve a minimum spectral efficiency $\Gamma^{\min}$; otherwise, a rate-violation penalty is applied. 
At decision epoch~$t$, the reward is defined as
\[
r_t = -\frac{\sum_{l=1}^{L} n_{l,t}}{L N}
      - \lambda_t\, R_{\mathrm{vio}}(t)
      - \zeta_t ,
\]
where $\lambda_t>0$ adaptively balances energy efficiency and reliability through an online Lagrangian update 
$\lambda_{t+1}=\max(0,\,\lambda_t+\eta(\bar{R}_{\mathrm{vio}}-R^\ast))$. 
The term $\zeta_t \ge 0$ represents the infeasibility penalty applied when the agent 
proposes invalid actions such as deactivating all APs or exceeding physical limits. 
In implementation, $\zeta_t$ combines a fixed penalty of $1.0$ for total shutdowns 
and a small incremental penalty of $0.05$ for each invalid action, thus ensuring that 
the agent maintains feasible activation patterns during training.
This unified reward formulation embeds the spectral efficiency constraint in \eqref{eq:power_min_opt0:SINR_constraint} directly into the learning process, 
enabling stable and power-aware near-real-time control.

\subsection{Algorithm Design}
We employ the Proximal Policy Optimization (PPO) algorithm, 
a state-of-the-art on-policy method that efficiently handles multi-dimensional discrete action spaces~\cite{schulman2017proximal}. 
PPO provides stable policy updates and, in our preliminary tests, achieved faster convergence and lower variance than 
value-based alternatives, confirming its suitability for near-real-time O-RU and antenna control.

During inference, the trained PPO agent takes actions for given UE channel gains, without requiring full CSI. Two initialization modes are considered:  
\texttt{open}, with all antennas active, and \texttt{close}, with all antennas inactive. The solution with the least SE requirement violation is selected. If both solutions satisfy SE requirements, the lowest power-consuming solution is selected. 
To promote further antenna deactivation without creating SE requirement violations, a greedy refinement stage is introduced to prune redundant antenna activations 
identified by the RL policy. The inference and refinement process is summarized in Algorithm~\ref{alg:inference_greedy}. \texttt{Simulate}$(\mathbf{n}')$ in the algorithm evaluates $\mathbf{n}'$ and returns $(P', R_{\mathrm{vio}}(t))$ by randomly creating small-scale fading channels with given channel gains.

\subsection{Heuristic Benchmark}
To benchmark the proposed algorithm, we design a lightweight heuristic antenna allocation algorithm. The guiding principle is to activate more antennas at  O-RU--UE  pairs with stronger large-scale channel conditions. Given the dB level large-scale fading coefficients 
$\beta^{\mathrm{dB}}_{l,k}$, the algorithm first weights among UEs, where the weight of UE~$k$ becomes
  $
        v_k = \frac{\|\boldsymbol{\beta}^{\mathrm{dB}}_k\|_\alpha}
                    {\sum_{k'=1}^K \|\boldsymbol{\beta}^{\mathrm{dB}}_{k'}\|_\alpha}.
   $, and $\|\boldsymbol{\beta}^{\mathrm{dB}}_{k'}\|_\alpha$ denotes the $\alpha$-norm of the channel gain.  
Then, for each UE $k$, the relative contribution of O-RU $l$ is weighted by
 $\tilde{v}_{l,k}  = \frac{\beta^{\mathrm{dB}}_{l,k}}{\sum_{l'=1}^L \beta_{l',k}}$. The activated antennas at O-RU $l$ is decided by 
 \begin{equation}
     n_{l} = \min\!\bigl\{N,\;\lfloor \sum_{k=1}^K v_k \tilde{v}_{l,k} LN \rfloor\bigr\},
 \end{equation}
ensuring that each  O-RU does not exceed $N$.

\begin{figure}[!t]
\vspace*{0.1in}
\centering
\begin{minipage}{0.95\columnwidth}
    \begin{algorithm}[H]
\caption{EARL: Energy-Aware Adaptive Antenna Control with Reinforcement Learning}
\label{alg:inference_greedy}

\begin{algorithmic}[1]

\State \textbf{Input:} Trained PPO policy $\pi_\theta$, environment $\mathcal{E}$, max steps $T$, $\boldsymbol{\Phi}$
\State $\mathcal{C}\gets\emptyset$

\For{mode $\in$ \{\texttt{close}, \texttt{open}\}}
    \State $s_0 \gets \mathcal{E}.\texttt{reset}(\text{mode})$
    \State  Extract initial activation $\mathbf{n}_0$ from $s_0$
    \State $\mathbf{n}\gets\mathbf{n}_0$
    \For{$t=1$ to $T$}
        \State $\mathbf{a}_t \gets \pi_\theta(\boldsymbol{\Phi},\mathbf{n})$
        \State Update activation $\mathbf{n}\gets\text{Clip}(\mathbf{n}+\Delta(\mathbf{a}_t))$
        \State Form next state $s_t \gets (\boldsymbol{\Phi},\mathbf{n})$
    \EndFor
    \State Evaluate final configuration: 
    \State $(P,R_{\mathrm{vio}})\gets\texttt{Simulate}(\mathbf{n})$
    \State $\mathcal{C}\gets\mathcal{C}\cup\{(\mathbf{n},P,R_{\mathrm{vio}})\}$
\EndFor

\State Select $(\mathbf{n},P,R_{\mathrm{vio}})\in\mathcal{C}$ by priority: lowest outage, then lowest power
\State \Comment{Greedy refinement:}
\For{each O-RU $l=1$ to $L$}
    \While{$n_l>0$}
        \State $\mathbf{n}'\gets\mathbf{n}$ with $n'_l\gets n_l-1$
        \If{$R'_{\mathrm{vio}}>R_{\mathrm{vio}}$ }
            \State \textbf{break}
        \Else
            \State $\mathbf{n}\gets\mathbf{n}'$;\; $P\gets P'$;\; $R_{\mathrm{vio}}\gets R'_{\mathrm{vio}}$
        \EndIf
    \EndWhile
\EndFor

\State \textbf{Output:} Final activation $\mathbf{n}^{\text{final}}=\mathbf{n}$,  $P_{\mathrm{tot}}$, $R_{\mathrm{vio}}$.

\end{algorithmic}
\end{algorithm}

\end{minipage}
\end{figure}

\section{Numerical Analysis}

This section presents the simulation framework to evaluate the effectiveness of EARL in cell-free massive MIMO networks. Network simulations are conducted in MATLAB, while the RL agent is trained in Python.

We consider a square coverage area of $400 \text{m} \times 400 \text{m} $ containing 
$L = 16$ O-RUs arranged in a regular grid. 
UEs are uniformly and independently distributed within the area. 
The main system configuration, power allocation parameters, and power consumption constants are summarized in Table~\ref{tab:sim_and_hyper}.

\begin{table}[t]
\centering
\caption{Simulation Parameters}
\label{tab:sim_and_hyper}
\begin{tabular}{lclc}
\toprule
\multicolumn{4}{l}{\textit{System Configuration}} \\
\midrule
\textbf{Parameter} & \textbf{Value} & \textbf{Parameter} & \textbf{Value} \\
$\tau_c$, $\tau_p$ & 192, 6 & $N$ & 8  \\
$B_W$, $f_s$ & 20, 30.72\,MHz & ASD & $15^\circ$  \\
$N_{\mathrm{DFT}}$, $N_{\mathrm{used}}$ & 2048, 1200 & $T_s$ & 71.4\,$\mu$s   \\
  $p$, $\rho_{\text{tot}}$ & 100, 200\,mW &  $\sigma_{\mathrm{cool}}$& 0.9  \\
  $(\upsilon,\kappa)$ & $(-0.5,\,0.5)$ & $\Delta_{\mathrm{Tr}}$ & 4  \\
$P_{\mathrm{RU,sta}}$ & 6.8\,W per ant. & $P_{\mathrm{proc}}$ & 74\,W \\
$P_{\mathrm{idle}}$, $P_{\mathrm{fixed}}$ & 20.8\,W, 120\,W & $C^{\max}_{\mathrm{RU}},C^{\max}_{\mathrm{cloud}}$ & 360\,GOPS \\
$P_{\mathrm{opt}}$, $P_{\mathrm{OLT}}$  & 1.8\,W, 20\,W & $\Delta_{\mathrm{ptp}}$ & 4.6 \\
\midrule
\multicolumn{4}{l}{\textit{PPO Training Hyperparameters}} \\
\midrule
Learn. rate & $10^{-4}$ & Discount fact. & 0.999 \\
GAE par. & 0.97 & Clip range $\epsilon$ & 0.2 \\
Batch size & 256 & Target KL & 0.025 \\
Netw. struct. & [256, 256, 256] & $\lambda_0$, $\eta$  & 0.3,  0.05 \\
\bottomrule
\end{tabular}
\end{table}

The proposed EARL algorithm is trained using the PPO algorithm with the hyperparameters listed in Table~\ref{tab:sim_and_hyper}. 
Each episode corresponds to a distinct channel realization with independently generated UE positions\footnote{In real-time implementation, the small-scale channel coefficients are not required. The algorithm only needs the channel gain of all O-RUs and UEs. Random samples can be generated from the channel distribution, and the expected SE can be approximated by the average of many samples.}. 
Feature normalization is applied to all state variables to improve numerical stability. 
Training is terminated using an early-stopping criterion when the mean reward variation over ten consecutive epochs falls below $10^{-3}$, ensuring efficient convergence without overfitting. We set $\alpha = 1$ for the heuristic benchmark algorithm since it provided the best performance.

\begin{figure}[tb]
\vspace*{0.1in}
    \centering
    \includegraphics[width=\linewidth]{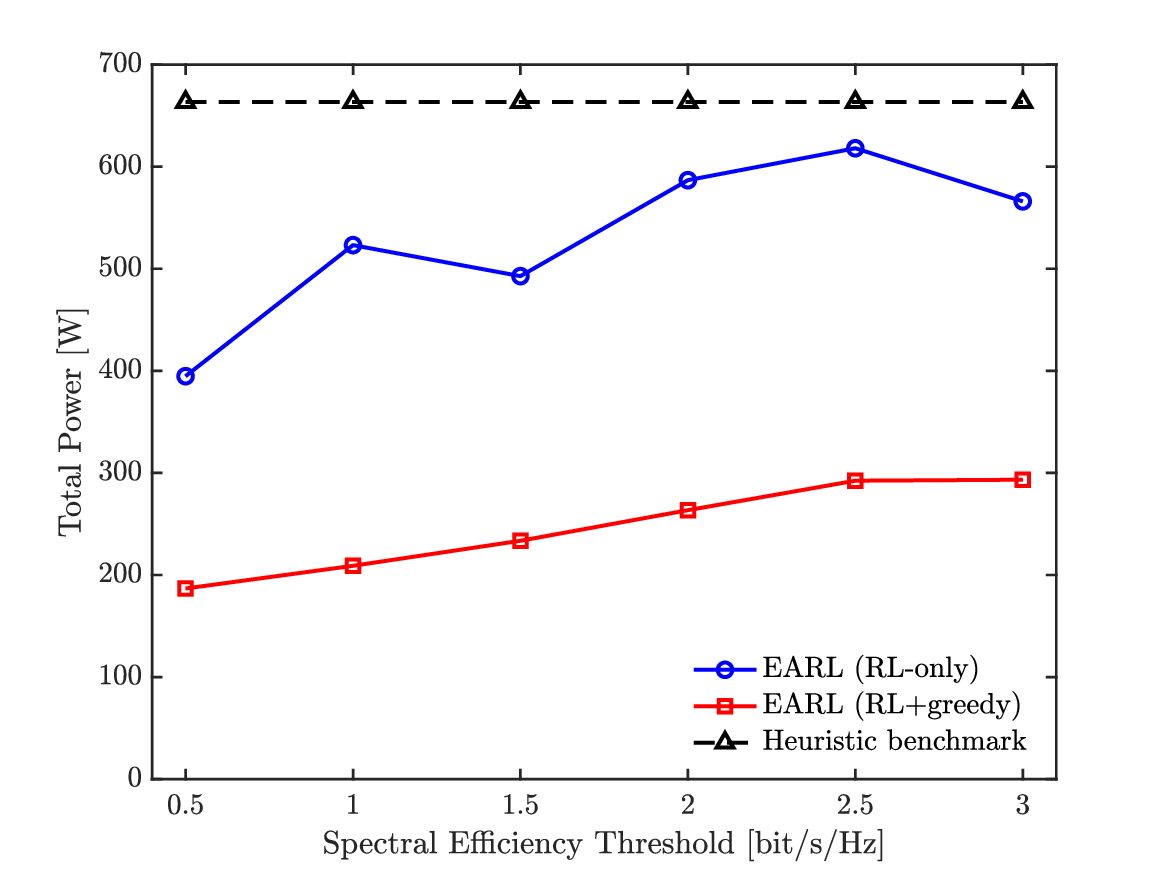}
    \caption{Average power consumption comparison under different SE thresholds.}
    \label{fig:greedy_comparison}
\end{figure}

Fig.~\ref{fig:greedy_comparison} illustrates the total power consumption for different SE thresholds for $K=4$ UEs. The heuristic algorithm lowers the power consumption by half. However, the power consumption does not scale with the SE requirements, since the heuristic algorithm only considers the channel strengths for a scalable and fast operation. In contrast, the RL-based approach dynamically adapts to varying SE requirements, lowering the power consumption by $45\%$ compared to the heuristic benchmark. The fluctuation in the RL-only algorithm demonstrates the approximation and generalization limits of the PPO algorithm. The proposed EARL algorithm combines RL decision with a greedy post-processing, and further reduces the power consumption by  $50\%$ compared to the RL-only benchmark. Even under the most stringent SE requirement, the proposed algorithm can reduce the total power consumption by $87\%$, thanks to the flexibility obtained by the centralized precoding capabilities.

\begin{table}[tb]
\centering
\caption{Performance comparison under $SE_{\text{thr}} = 1.5$.}
\begin{tabular}{lcccc}
\toprule
\textbf{Metric} & \textbf{UE = 4} & \textbf{UE = 6} & \textbf{UE = 8} \\
\midrule
\multicolumn{4}{c}{\textbf{EARL (RL-only)}} \\
\midrule
Total Power [W] & 493 & 624 & 986 \\
Average SE [bit/s/Hz] & 6.1 & 5.6 & 6.5 \\
\midrule
\multicolumn{4}{c}{\textbf{EARL (RL+greedy)}} \\
\midrule
Total Power [W] & 234 & 292 & 465 \\
Average SE [bit/s/Hz] & 2.9 & 3.0 & 3.6 \\
\bottomrule
\end{tabular}
\label{tab:greedy_se15}
\end{table}

Table~\ref{tab:greedy_se15} demonstrates that the RL-agent prioritizes the SE requirements of the users, still reducing the power consumption from $33\%$ to $70\%$ compared to fully activating all network elements. Due to this prioritization, the RL-agent also obtains a higher average SE, where greedy processing reduces this close to the SE threshold, lowering the power consumption by half of the RL-agent. While the average SE with greedy processing is still significantly higher than the SE threshold of UEs, further reduction in active antennas or O-RUs creates rate violations. Jointly allocating transmit power with the RL-agent can lower the average SE to the SE threshold limits, but will only provide marginal gains to the total power consumption~\cite{ozan_asilomar}. 

\begin{table}[t]
    \centering
    \caption{Average run-time comparison of different methods considering $K=4$ under clock speed $3.2$ GHz.}
    \label{tab:mean_second}
    \begin{tabular}{l c}
        \hline
        \textbf{Method} & \textbf{Average Run-Time (s)} \\ 
        \hline
        EARL (RL-only)          & 0.22 \\
        EARL (RL+greedy)     & 1.98 \\
        Heuristic    & 0.07 \\
        \hline
    \end{tabular}
\end{table}
It can be observed from Table~\ref{tab:mean_second} that the heuristic baseline 
achieves the shortest run-time, 
requiring only a few arithmetic operations.
In contrast, the RL-only agent has a higher run-time due to the size of the PPO agent, but still significantly below the 1-2 seconds limit for near-RT xApp. Finally, the greedy postprocessing increases the run-time significantly, reaching the control time limit. Furthermore, as the size of the network increases, the run-time of the greedy refinement will also increase, making it a practical refinement solution for the smaller networks.

\begin{figure}[t]
    \centering
    \includegraphics[width=\linewidth]{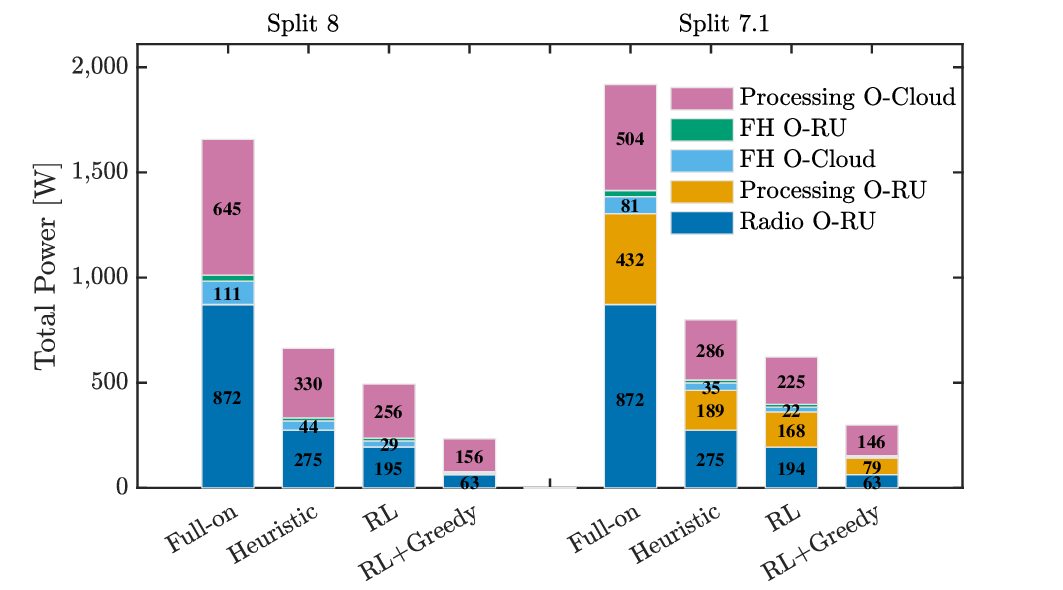}
    \caption{Power consumption breakdown under Split~8 and Split~7.1. }
    \label{fig:energy_breakdown_unified}
\end{figure}

Fig. \ref{fig:energy_breakdown_unified} compares total power consumption for functional splits 8 and 7.1. The key difference lies in processing distribution: in Split 7.1, the O-Cloud transmits frequency-domain signals to O-RUs, and each O-RU performs inverse fast Fourier transform (IFFT), consuming both idle processing and antenna-dependent processing power in O-RUs. In contrast, Split 8 centralizes all processing in the O-Cloud, enabling processing load sharing. Although O-Cloud power usage rises, overall network energy consumption decreases by $30\%$ due to efficient resource sharing. Thus, centralizing processing with Split 8 and deploying dense O-RUs improves spectral efficiency through centralized precoding and reduces total energy consumption through cloud-based processing sharing.

\section{Conclusion}
In this work, we proposed an energy-aware adaptive antenna control algorithm with reinforcement learning (EARL) in O-RAN cell-free massive MIMO networks with centralized precoding. The proposed algorithm minimizes the total power consumption in the network, while ensuring the spectral efficiency requirements of the user equipments (UEs) are not violated. 
The numerical results demonstrate that the power consumption can be reduced by $81\%$ and $50\%$ compared to the full-on and heuristic benchmarks, respectively. The RL-only version operates in $220$ ms, well within the O-RAN near-real-time limit of $2$ s, while the greedy-refinement step increases runtime to 2 s but further cuts power consumption by $50\%$. These results demonstrate EARL’s energy-efficient and fast operation in O-RAN cell-free massive MIMO networks. Future work includes joint O-RU antenna activation and O-RU and UE association, and improving the scalability of the algorithm.

\bibliographystyle{IEEEtran}
\bibliography{IEEEabrv,refs}

\end{document}